# On the relationship between the diffuse reflection and bounce-back boundary condition in the continuum limit

Jianping Meng, Xiao-Jun Gu, and David R Emerson

Scientific Computing Department, STFC Daresbury Laboratory, Warrington WA4 4AD, United Kingdom

Yong Peng and Jianmin Zhang

State Key Laboratory of Hydraulics and Mountain River Engineering, Sichuan University, Chengdu, 610065, P. R. China

**Abstract:**
In this work, we show that the widely used bounce-back boundary condition is an incomplete form of the diffuse reflection boundary condition at the continuum limit for lattice Boltzmann simulations. By utilizing this fact, we can force the diffuse reflection scheme to work at its continuum limit so that the no-slip boundary condition can be implemented without any non-physical slip velocity error being induced by the standard bounce-back scheme. The revised boundary formulation is validated numerically by solving a pressure-driven channel flow, a lid-driven cavity flow and channel flow around a square cylinder.

## 1. Introduction

For lattice Boltzmann (LB) simulations, it is of great importance to correctly implement the no-slip wall boundary condition for a wide range of flow problems. The bounce-back (BB) scheme is perhaps the most widely used approach among various implementations. It is easy to apply and works particularly well with the cut-cell mesh technique, i.e., a series of zig-zag stairs are used to approximate any complex boundary curve or surface. A moving boundary can also be treated via a simple modification as suggested in [1]. However, the BB scheme can suffer from a the non-physical slip velocity begin generated at the wall, which is discussed in detail in [2,3]. This can induce simulation errors in flows with both simple [4] and complex geometry, e.g., predicting permeability of a porous medium [5,6]. In [7], it is found that the BB scheme does not specify certain discrete velocities, and this manifests in the creation of the non-physical slip velocity. To correctly implement the no-slip velocity boundary, it is important to supplement the missing information. This is fairly easy for simple geometries but remains open for general flow problems. On the other hand, it is also worth noting that the BB scheme provides a complete definition if using a lattice without a zero velocity component and the cut-cell mesh technique, e.g., the D2Q4 lattice structure analysed in [7]. However, lattices with a zero velocity component are more common in the literature [8,9].

The diffuse reflection (DR) boundary scheme (also known as the kinetic boundary scheme) is believed to be consistent with physical effects for modelling slip velocity at the wall [10]. Therefore, it is often employed together with the discrete velocity model (DVM) for rarefied gas flows (see e.g., [11]). As the LB method may be considered as a special form of DVM, the DR scheme was naturally introduced into LB simulations [12]. The boundary scheme has not only been used to capture the rarefied gas effects [13,14], but it has also demonstrated its effectiveness in simulating a lid-driven cavity flows with high Reynolds ($Re$) number (upto 7500), and the resulting slip velocity shows negligible impact on typical simulations [15]. This arises because that the slip velocity become negligible with increasing Reynolds number and decreasing Knudsen ($Kn$) number. In the limit of $Kn \to 0$, the no-slip boundary will be fully recovered.

In this work, by investigating the limiting behavior of the DR scheme as $Kn \to 0$, we find that the BB scheme represents an incomplete form of the continuum limit of the DR scheme. Therefore, we are able to supplement the information missed by the BB scheme, and correctly implement the no-slip velocity boundary for a general problem. In the following, we will briefly introduce the lattice Boltzmann model and the two boundary conditions and then analyze the relationship between the two schemes. Based on the identified relationship, we devise an equilibrium diffuse reflection (EDR) boundary scheme for correctly implementing the no-slip boundary condition. Finally, we validate the proposed boundary formulation by simulating a pressure-driven channel flow, a lid-driven cavity flow and channel flow around a square cylinder.

## 2. Lattice Boltzmann model

The lattice Boltzmann model can be derived via various procedures, see, e.g. [16–18], and may be considered a special form of DVM [13,19]. Its governing equation can be written as

$$\frac{\partial f_\alpha}{\partial t} + \boldsymbol{C}_\alpha \cdot \frac{\partial f}{\partial \boldsymbol{r}} = \frac{1}{\tau}(f_\alpha^{eq} - f_\alpha), \tag{1}$$

which describes the evolution of the single-particle distribution function $f_\alpha(\boldsymbol{r},t)$ for the αth discrete velocity $\boldsymbol{C}_\alpha = (c_x, c_y)$ at position $\boldsymbol{r} = (x,y)$ for two-dimensional flows. For simplicity, the particle interaction can be modeled using a relaxation term towards the discrete equilibrium distribution function $f_\alpha^{eq}(\boldsymbol{r},t)$. For incompressible and isothermal flows, the Maxwellian distribution can often be truncated into a second-order polynomial, i.e.,

$$f_\alpha^{eq} = \rho g_\alpha^{eq} = \rho w_\alpha \left\{ 1 + \frac{\boldsymbol{C}_\alpha \cdot \boldsymbol{V}}{\theta_0} + \frac{1}{2}\left[\frac{(\boldsymbol{C}_\alpha \cdot \boldsymbol{V})^2}{\theta_0^2} - \frac{\boldsymbol{V} \cdot \boldsymbol{V}}{\theta_0}\right]\right\}, \tag{2}$$

which is determined by the density, $\rho$, the fluid velocity, $\boldsymbol{V} = (u,v)$, and the reference temperature, $\theta_0$. The relaxation time, $\tau$, is related to the viscosity, $\mu$, via the relation $\mu = p\tau$. The Reynolds number and Knudsen number can also be defined accordingly, see [15] for details. Together with Eq. (2), a D2Q9 lattice model with nine discrete velocities

$$c_x = \sqrt{3\theta_0}(0, 1, 0, -1, 0, 1, -1, -1, 1), \tag{3}$$

and

$$c_y = \sqrt{3\theta_0}(0, 0, 1, 0, -1, 1, 1, -1, -1), \tag{4}$$

is often adopted, where the corresponding weights are

$$w = \left(\frac{4}{9}, \frac{1}{9}, \frac{1}{9}, \frac{1}{9}, \frac{1}{9}, \frac{1}{36}, \frac{1}{36}, \frac{1}{36}, \frac{1}{36}\right). \tag{5}$$

With the discretized model, the macroscopic fluid density and flow velocity can be obtained from

$$\rho = \sum_{\alpha=0}^{8} f_\alpha \qquad \rho \boldsymbol{V} = \sum_{\alpha=0}^{8} f_\alpha \boldsymbol{C}_\alpha. \tag{6}$$

Eq. (1) can be solved numerically by using the scheme

$$f_\alpha(t+dt, \boldsymbol{r}+\boldsymbol{c}_\alpha dt) - f_\alpha(t,\boldsymbol{r}) = \frac{dt}{\tau + 0.5\, dt}(f_\alpha^{eq} - f_\alpha), \tag{7}$$

which is second-order accurate in both time and space [20].

## 3. Continuum limit of diffuse-reflection boundary condition

For convenience, we briefly give the definition of the DR scheme and the BB scheme for the D2Q9 lattice.

With the DR scheme, an outgoing particle completely forgets its history and its velocity is re-normalized by the Maxwellian distribution. Moreover, we also assume that the effective particle-wall interaction time is small in comparison to any characteristic time of interest and no permanent adsorption occurs. To write down the mathematical description, the discrete velocities are categorized into three sets, i.e., incoming (blue solid lines, notated as $\xi_I$), outgoing (red dashed lines, notated as $\xi_O$), and parallel (black dash-dot lines, notated as $\xi_P$), see Figure 1 and Figure 2.

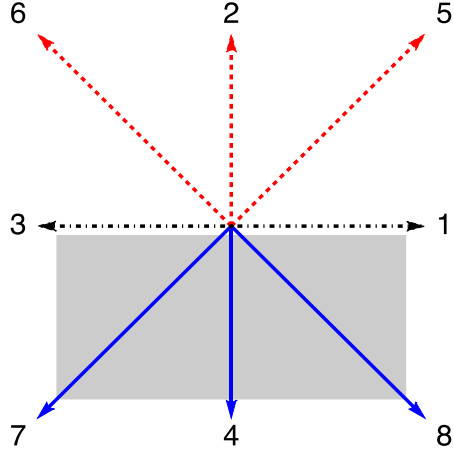

Figure 1 Schematic of a bottom wall boundary. Wall is illustrated by using gray shading.

For a planar surface, the surface normal $\boldsymbol{n}$ is often used to distinguish the three types of discrete velocities. As has been shown in Figure 1, for a bottom wall moving a velocity $\boldsymbol{V}_b = (u_b, v_b)$, we have $\alpha \in \xi_I$ if $(\boldsymbol{C}_\alpha - \boldsymbol{V}_b) \cdot (0,1) < 0$, $\beta \in \xi_O$ if $(\boldsymbol{C}_\beta - \boldsymbol{V}_b) \cdot (0.1) > 0$, and $\gamma \in \xi_P$ if $(\boldsymbol{C}_\gamma - \boldsymbol{V}_b) \cdot (0.1) = 0$. For generality, the vertical component $v_b$ is also considered which is often zero for the bottom wall shown in Figure 1. However, we note that a knowledge of the surface normal is not essential in the actual implementation. By using the cut-cell technique, it is possible to distinguish these three types by judging whether a particle will enter into the wall or not. Thus, the algorithm at boundary points can be specified according to the boundary types (e.g., the planar type shown in Figure 1 and the two corners shown in Figure 2).

For the D2Q9 lattice, the DR scheme can then be written as

$$f^b_{\alpha\in\{\xi_o,\xi_P\}} = \frac{\sum_{\beta\in\xi_I} f^b_\beta}{\sum_{\alpha\in\xi_O} g^{eq}_\alpha(\boldsymbol{V}_b)} g^{eq}_{\alpha\in\{\xi_o,\xi_P\}}(\boldsymbol{V}_b) \tag{8}$$

The first RHS term can be considered as the fluid density $\rho_b$ at the wall obtained according to mass conservation, i.e.,

$$\rho_b = \frac{\sum_{\beta\in\xi_I} f^b_\beta}{\sum_{\alpha\in\xi_O} g^{eq}_\alpha(\boldsymbol{V}_b)}. \tag{9}$$

At the bottom wall boundary, the set $\xi_I$ includes discrete velocities {7,4,8}, $\xi_O$ {6,2,5} and $\xi_P$ {0,1,3}. Due to the exact advection associated with the particle nature of the lattice Boltzmann algorithm, we can calculate the mass of particles that cross the boundary by using the distribution function instead of flux terms, see also [21].

It is straightforward to write down the formulations for other planar boundaries. Corners are slightly more complicated. For instance, as shown in Figure 2, discrete velocities {1,2,5} can be considered as the set $\xi_I$, {3,4,7} as the set $\xi_O$ and {0,6,8} as the set $\xi_P$. Thus, we can use Eq. (8) implement the boundary scheme.

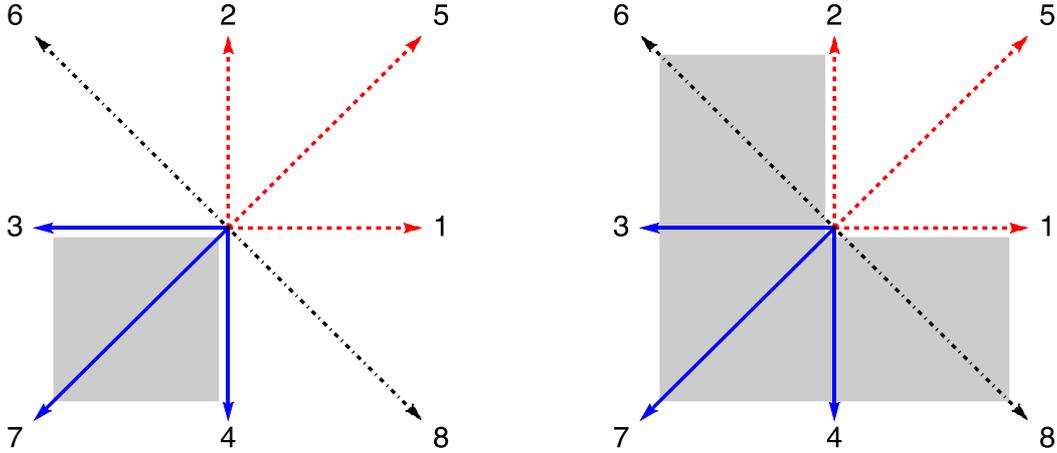

Figure 2 Schematic of two types of wall boundary corners. Wall is illustrated by using gray shading. The discrete velocities are categorized into incoming (blue solid lines), outgoing (red dashed lines), and parallel (black dash-dot lines).

The main idea of the BB scheme is that an incoming particle distribution impacting the wall bounces back into the fluid domain in the opposite direction, and therefore the stationary no-slip boundary is supposed to be enforced. Ladd [1] generalized the scheme for the moving body by using

$$f^b_\alpha = f^b_{\bar{\alpha}} + 2w_\alpha \rho_b \frac{\boldsymbol{C}_\alpha \cdot \boldsymbol{V}_b}{\theta_0}, \tag{10}$$

which relates the unknown incoming $f^b_\alpha$ to its known counterpart $f^b_{\bar{\alpha}}$ with opposite discrete velocity (i.e., $\boldsymbol{C}_{\bar{\alpha}} = -\boldsymbol{C}_\alpha$) and a simple modification for specifying the boundary velocity. For the specific D2Q9 lattice and the bottom wall boundary, we can accordingly treat the discrete velocity pair from the set $\xi_O$ and the set $\xi_I$, respectively. However, as discussed in [7], there is no definition for those in the set $\xi_P$.

It is known that the DR boundary scheme will produce a slip velocity at the wall with finite Knudsen number [22]. At the limit of $Kn \to 0$, the slip velocity will naturally vanish, which can be seen from the slip velocity boundary formula. Therefore, according to Eq. (8), $f_\alpha^b = \rho_b g_\alpha^{eq}(V_b)$ for $\alpha \in \{\xi_O, \xi_P\}$. Moreover, at the limit, the fluid is always in local equilibrium, i.e., $f_{\beta \in \xi_I}^b = \rho_b g_\beta^{eq}(V_b)$ for $\beta \in \xi_I$. We note that two sets of equilibrium distribution must adopt the same density since discontinuities must be removed by the very fast particle collisions. Noticing the fact that for a pair of discrete velocities $C_\alpha$ and $C_{\bar\alpha} = -C_\alpha$, we have

$$\rho_b g_\alpha^{eq}(V_b) = \rho_b g_{\bar\alpha}^{eq}(V_b) + 2w_\alpha \rho_b \frac{C_\alpha \cdot V_b}{\theta_0}, \tag{11}$$

which can be obtained by substituting velocities $C_\alpha$ and $C_{\bar\alpha}$ into Eq. (2). Thus, we immediately obtain the relation presented in Eq. (10), and we can conclude that, at the limit of $Kn \to 0$, the DR scheme reduces to the BB scheme for discrete velocity pairs in the sets $\xi_I$ and $\xi_O$.

The foregoing observation presents an interesting foundation for the BB boundary condition, in particular for a moving boundary. That is, the BB scheme assembles the limiting behavior of the DR scheme. This observation is also consistent with the finding in [6], the non-physical slip velocity will reduce with decreasing the Knudsen number since the distribution functions for discrete velocities in the set $\xi_P$ will tend to their equilibrium and will naturally cancel each other.

To remove the slip velocity, it is now possible to ensure the DR scheme operating under its continuum limit (i.e., $Kn \to 0$). For a pair of $C_\alpha$ and $C_{\bar\alpha}$, we directly apply Eq. (10) rather than the Eq. (8)

$$f_\alpha^b = f_{\bar\alpha}^b + 2w_\alpha \rho_b \frac{C_\alpha \cdot V_b}{\theta_0} \text{ for } \alpha \in \xi_O \text{ and } \bar\alpha \in \xi_I, \tag{12}$$

and, for the discrete velocity in the set $\xi_P$, we apply

$$f_{\alpha \in \xi_P}^b = \rho_b g_{\alpha \in \xi_P}^{eq}(V_b). \tag{13}$$

Thus, we can eliminate the slip velocity, and we call Eqs. (12) and (13) 'equilibrium diffuse reflection' (EDR) scheme since the DR scheme is operating under the complete equilibrium state. In comparison to the BB scheme, the Eq. (13) helps to eliminate the ambiguous definition for the discrete velocities in the set $\xi_P$. Using Eqs (12) and (13), it is easy to verify that

$$\sum_{\alpha \in \xi_O} f_\alpha^b C_\alpha + \sum_{\beta \in \xi_I} f_\beta^b C_\beta + \sum_{\gamma \in \xi_P} f_\gamma^b C_\gamma \equiv \rho_b V_b,$$

so that the no-slip boundary condition is exactly implemented.

We also need to alter Eq. (9) of $\rho_b$ for the EDR scheme. It is inconsistent with Eq. (12) and may produce density discontinuities at solid boundary. To avoid this situation, we first write the local mass conservation as

$$\sum_{\alpha \in \xi_O} f_\alpha^b + \sum_{\beta \in \xi_I} f_\beta^b + \sum_{\gamma \in \xi_P} f_\gamma^b = \rho_b. \tag{14}$$

Substituting Eqs. (12) and (13) into Eq. (14), the only unknown quantity is the density, $\rho_b$, which can be obtained by solving the equation. For instance, the formulations at the boundary points shown in Figure 1 and Figure 2 can be calculated as

$$\rho_b = \frac{6(f_4 + f_7 + f_8)\theta_0}{\theta_0 + v_b^2 - \sqrt{3\theta_0}v_b},$$

and

$$\rho_b = \frac{36(f_3 + f_4 + f_7)\theta_0}{9\theta_0 + 3u_b^2 - 5\sqrt{3\theta_0}u_b + 3u_b v_b + 3v_b^2 - 5\sqrt{3\theta_0}v_b},$$

respectively. The other planar boundaries and boundary corners can be treated in an analogous way. In these discussions, we are using physical units so that the reference temperature $\theta_0$ is always presented. For the transformation to lattice units, one may refer to [15] for detail.

## 4. Numerical validations for equilibrium diffuse reflection scheme

In the following, we will conduct a series of numerical validations for three typical benchmark problems. The first benchmark is the classical two-dimensional pressure-driven channel flow where the top wall is in motion. The channel length and height are set to be $L$ and $H$, respectively, and we set the ratio $L/H = 100$. The pressure difference is set to be $0.00005 \, \rho_0 \theta_0$ and the pressure at the outlet is set to be $\rho_0 \theta_0$. The fluid viscosity is $0.0017 \, \rho_0 \sqrt{\theta_0} \, H$. The velocity of the top wall is $(0.00001 \sqrt{\theta_0}, 0)$. For this configuration, the density at the outlet, $\rho_0$, is chosen as the reference density and the wall temperature, $\theta_0$, as the reference temperature. The corresponding Knudsen number is 0.0017 using the channel height as the reference length. In [15], there is a detailed discussion on the transformation between various physical or lattice units, which, for brevity, is not discussed here.

The pressure profiles at the inlet and outlet including the corner points are specified using a first-order extrapolation scheme which is implemented in the way described in, e.g., [23]. The wall boundary is enforced in the way described by Eqs. (12) and (13).

The numerical results (see the left part of Figure 3) agree well with the analytical solution of the NS equations, even using a mesh of $1000 \times 10$ cells. It is also observed that there is no non-physical slip velocity at the wall boundary nodes.

The numerical accuracy is measured for the stream-wise velocity component along the vertical centerline of the channel by conducting four simulations with 5, 10, 20, and 40 cells in the vertical ($Y$) direction. The errors are calculated against the analytical solution by using the $L^2$ norm for a vector $\boldsymbol{\phi}$, i.e.,

$$E = \frac{\|\boldsymbol{\phi} - \boldsymbol{\phi}_R\|_2}{\|\boldsymbol{\phi}_R\|_2}. \tag{15}$$

where $\|\boldsymbol{\phi}\|_2 = \sqrt{\phi_1^2 + \cdots \phi_n^2}$ and the reference solution is denoted by $\boldsymbol{\phi}_R$. In this case, we consider the velocities at the six grid points specified by the simulations with five cells. The results in the right part of Figure 3 shows an accuracy slightly better than second-order. Since there is no slip velocity at the wall boundary, no error is introduced into the bulk solution and we would expect to achieve the second-order accuracy of Eq. (7).

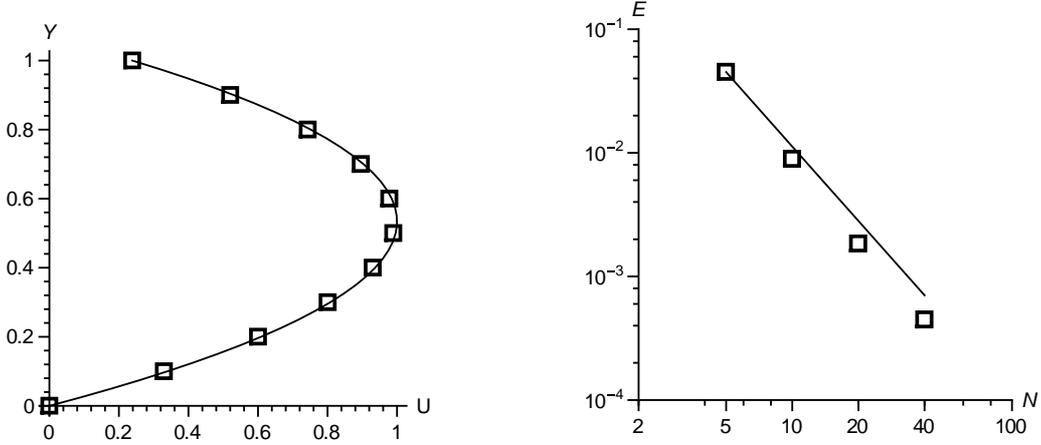

Figure 3 (Left) Profiles of stream-wise velocity component. The analytical solution is shown as the solid line. The speed is normalized by its maximum; (Right) Numerical accuracy. The ideal slope (-2) is shown using the solid line.

The second case is the lid-driven cavity flow, where the lid speed is fixed at $0.1\sqrt{\theta_0}$ and the wall temperature is held at the reference temperature $\theta_0$. To validate the boundary condition, we simulate three $Re$ numbers i.e., 100, 400 and 1000. By using $256 \times 256$ cells, we compare the solutions with those in [24]. As has been shown in Figure 4, the results agree well with each other, and importantly, there is again no non-physical slip velocity at the wall boundaries.

For the last validation, we consider a channel flow around a square cylinder at $Re = 25$ defined by the square width $D$ and the average velocity at the inlet. The channel length is set to be $L = 15D$ and height $H = 2.5D$. The left bottom corner of the square is located at the point $(4D, 0.75D)$, see Figure 6. At the inlet, we enforce a uniform velocity profile $(0.05\sqrt{\theta_0}, 0)$ by using the proposed EDR formulation. For the outlet, we use the first-order extrapolation scheme to set a uniform pressure (density) profile $\rho_0$, i.e., the reference density. Therefore, the viscosity is $0.002\,\rho_0\sqrt{\theta_0}\,D$ and the Knudsen number is $0.002$.

The simulation is conducted by using $900 \times 150$ cells, and the results are compared to those given by an in-house multi-block Navier-Stokes (NS) solver, THOR, in which the finite volume approach has been employed. The diffusive and source terms are discretized by a central difference scheme and the QUICK scheme [25] is employed for the convective terms. A collocated grid arrangement is used and the interpolation scheme of Rhie and Chow [26] is employed to eliminate any non-physical pressure oscillations. For the purpose of comparison, we use the same mesh for both solvers. In Figure 5, it is found that both solvers agree well with each other. In both corners and planar walls, the zero speed is correctly enforced. In Figure 6, it shows that the flow features are correctly captured. In the LB predictions, the stagnation point is located at $5.959D$ while the NS solver gives $5.947D$, the relative error is within $1.3\%$.

We also test the stability for this flow by using $60 \times 10$ cells, which means only 4 cells are distributed on the square. The Reynolds number is adjusted while the inlet velocity is set to be $(0.005\sqrt{\theta_0}, 0)$ and the outlet pressure is also $\rho_0\theta_0$. It is found that a stable simulation can be maintained where $Re = 250$, $Kn = 0.00002$ and the relaxation time in lattice units is 0.500866, which demonstrates satisfactory stability. Here a "stable" simulation means that the simulation lasts at least 300000 time steps without the occurrence of "*NAN*" and afterwards we stop monitoring.

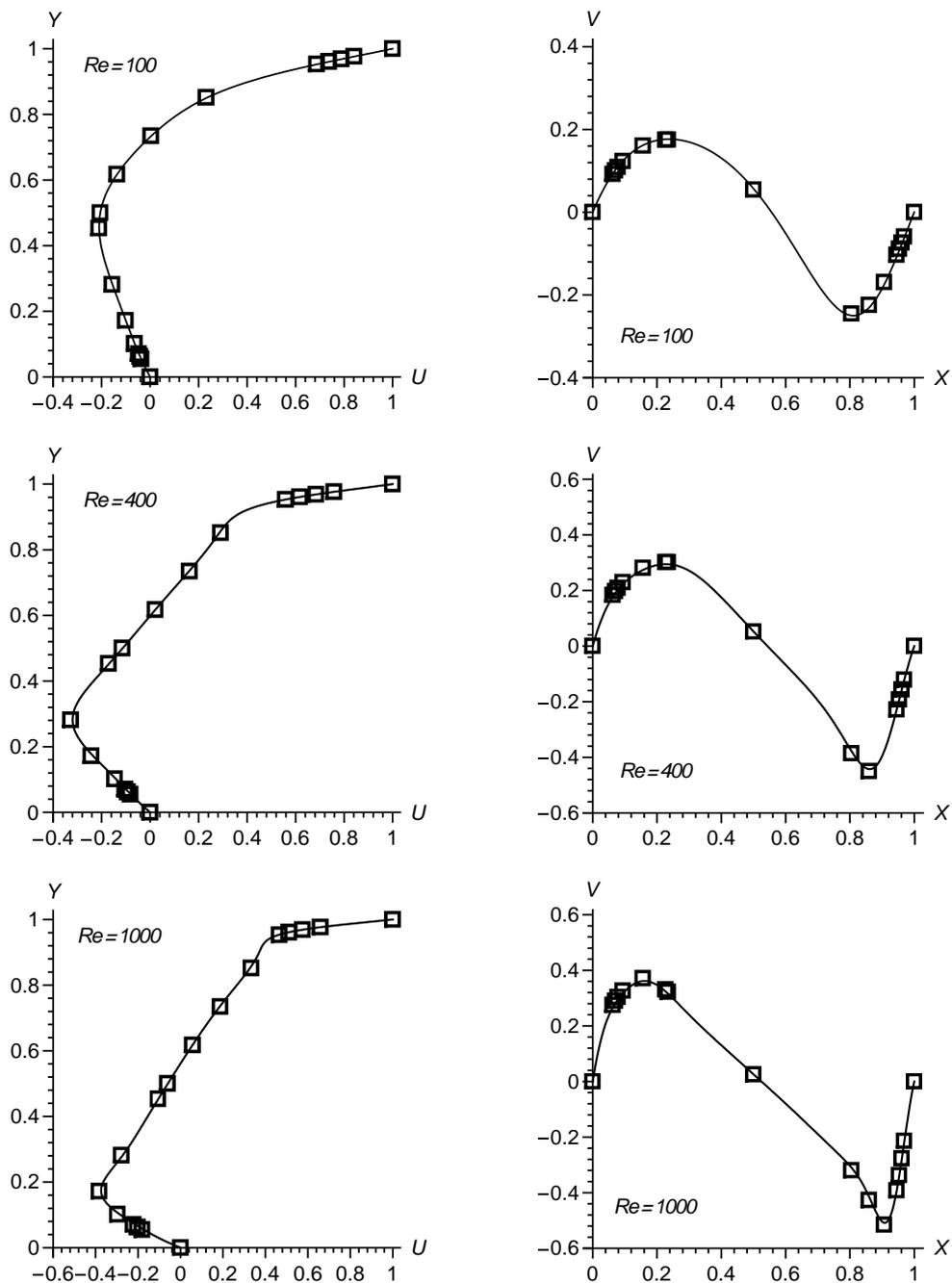

Figure 4 Profiles of horizontal (vertical) velocity component through the vertical (horizontal) centerline of the cavity. The horizontal and vertical components are normalized by the lid speed, which is set to be $0.1\sqrt{\theta_0}$ for all three cases. The benchmark solutions are taken from Ghia et al. [24].

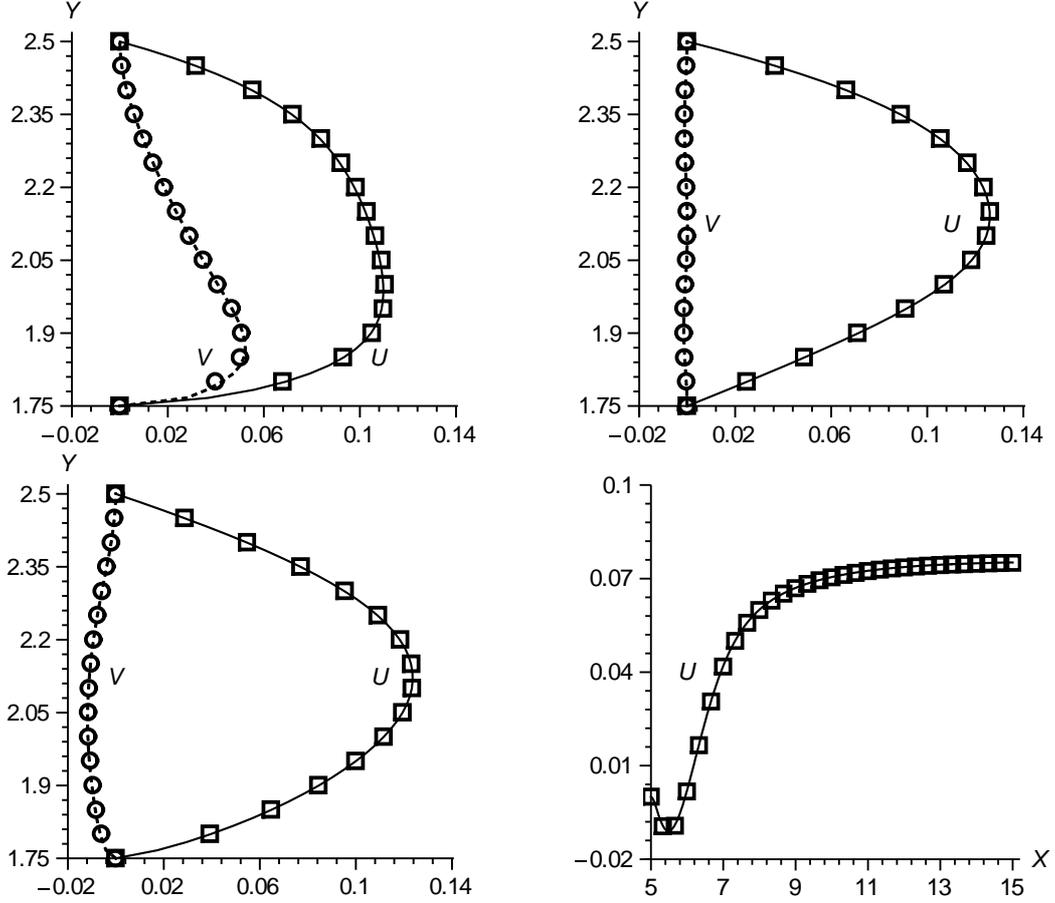

Figure 5 Velocity profiles along: (Top left) the vertical line from the top left corner of the square to the top wall of the channel; (Top right) the vertical line from the middle point of top wall of the square to the top wall of the channel; (Bottom left) the vertical line from the top right corner of the square to the top wall of channel; (Bottom right) the horizontal line of the channel after the square. The symbols are the results of LB simulations with the proposed boundary conditions.

## 4. Concluding remarks

By analyzing the limiting behavior of the DR boundary scheme at $Kn \to 0$, we have found that the commonly used BB boundary scheme is an incomplete form of the DR scheme, where the missing definition for the discrete velocities in the set $\xi_P$ induces the well-known non-physical slip velocity [7]. Utilizing this fact, we suggest the EDR scheme to implement the no-slip boundary condition, which ensures the DR scheme working under its equilibrium limit. To validate the scheme, numerical simulations are conducted for the pressure-driven channel flow, the lid-driven cavity flow, and the channel flow around a square cylinder, and satisfactory results are observed. We also note that the EDR scheme requires no extra effort in the implementation in comparison to the BB scheme although it could induce more computational overhead due to the calculation of the equilibrium function for the discrete velocities in the set $\xi_P$. Although the validations are conducted for two-dimensional flows with boundaries aligned to grid lines in this work, for avoiding error induced by the geometry, the EDR scheme, i.e., Eqs. (12), (13) and (14), can be applied to three-dimensional problems or curved boundaries since there are only minor modifications in comparison to the BB scheme, i.e., Eq. (13). In fact, we have presented treatments for the corners shown in Figure 2, which are ready for two-dimensional curved boundaries. Nevertheless, directly applying the scheme to zig-zag stairs will be subject to geometry error, and we will investigate possible ways of improving accuracy in the near future, e.g., combing with spatial interpolations [27].

**Acknowledgements**

Authors from the Daresbury Laboratory would like to thank the Engineering and Physical Science Research Council for their support under projects EP/N016602/1, EP/R029598/1, EP/P022243/1 and EP/N033841/1. Authors from the Sichuan University would like to thank the support from the National Natural Science Foundation of China (NSFC) (51579166) and the National Key Research and Development Program of China (2016YFC0401705). We also thank the Royal Society for their support on the bilateral visits under the project IE151119.

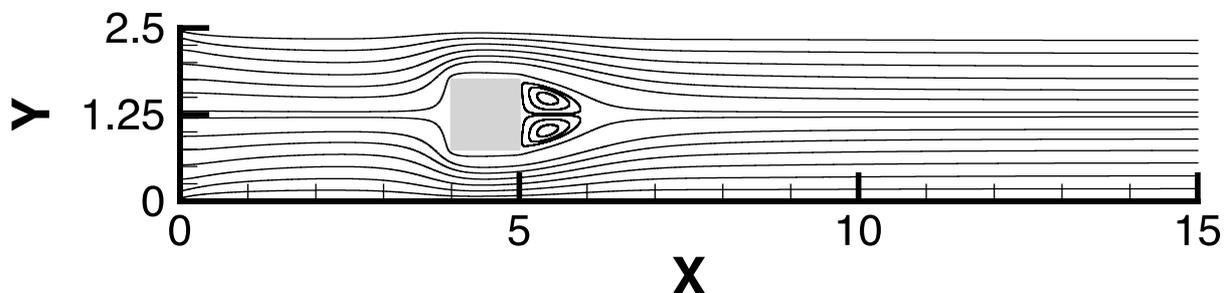

Figure 6 Streamline plot of the flow around a square cylinder at $Re = 25$.

**References**

[1]  A. J. C. Ladd, J. Fluid Mech. **271**, 285 (1994).
[2]  X. He, Q. Zou, L.-S. Luo, and M. Dembo, J. Stat. Phys. **87**, 115 (1997).
[3]  R. Cornubert, D. D'Humières, and D. Levermore, Phys. D **47**, 241 (1991).
[4]  L.-S. Luo, W. Liao, X. Chen, Y. Peng, and W. Zhang, Phys. Rev. E **83**, 1 (2011).
[5]  C. Pan, L.-S. Luo, and C. T. Miller, Comput. Fluids **35**, 898 (2006).
[6]  P. Prestininzi, A. Montessori, M. La Rocca, and S. Succi, Int. J. Mod. Phys. C **27**, 1650037 (2016).
[7]  J. Meng, X.-J. Gu, and D. R. Emerson, J. Comput. Sci. (2017).
[8]  S. Chikatamarla and I. Karlin, Phys. Rev. E **79**, 1 (2009).
[9]  X. Shan, Phys. Rev. E **81**, 1 (2010).
[10] J. Clerk Maxwell, Phil Trans R Soc Lond **170**, 231 (1879).
[11] R. Gatignol, Phys. Fluids **20**, 2022 (1977).
[12] S. Ansumali and I. V. Karlin, Phys. Rev. E **66**, 026311 (2002).
[13] J. Meng and Y. Zhang, J. Comput. Phys. **230**, 835 (2011).
[14] J. Meng, Y. Zhang, N. G. Hadjiconstantinou, G. A. Radtke, and X. Shan, J. Fluid Mech. **718**, 347 (2013).
[15] K. Hu, J. Meng, H. Zhang, X.-J. Gu, D. R. Emerson, and Y. Zhang, Comput. Fluids **156**, 1 (2017).
[16] Y. H. Qian, D. D'Humieres, and P. Lallemand, EPL Europhys. Lett. **17**, 479 (1992).
[17] X. He and L.-S. Luo, Phys Rev E **56**, 6811 (1997).

[18] S. Ansumali, I. Karlin, and H. Öttinger, EPL Europhys. Lett. **63**, 798 (2003).
[19] X. W. Shan, X. F. Yuan, and H. D. Chen, J. Fluid Mech. **550**, 413 (2006).
[20] X. He, S. Chen, and G. D. Doolen, J Comput Phys **146**, 282 (1998).
[21] J. Meng and Y. Zhang, J. Comput. Phys. **258**, 601 (2014).
[22] X.-J. Gu and D. R. Emerson, J. Fluid Mech. **636**, 177 (2009).
[23] F. Verhaeghe, L. Luo, and B. Blanpain, J. Comput. Phys. **228**, 147 (2009).
[24] U. Ghia, K. N. Ghia, and C. T. Shin, J. Comput. Phys. **48**, 387 (1982).
[25] B. P. Leonard, Comput. Methods Appl. Mech. Eng. **19**, 59 (1979).
[26] C. M. Rhie and W. L. Chow, AIAA J. **21**, 1525 (1983).
[27] M. Bouzidi, M. Firdaouss, and P. Lallemand, Phys. Fluids **13**, 3452 (2001).